\documentclass[prl,twocolumn,superscriptaddress,showpacs]{revtex4}
\usepackage{graphicx}
\makeatletter

\begin{document}
\title{Numerical Measurements of the Spectrum in Magnetohydrodynamic Turbulence}
\author{Joanne Mason}
\author{Fausto Cattaneo}
\affiliation{Department of Astronomy and Astrophysics, University of Chicago, 
5640 S. Ellis Ave, Chicago, IL 60637; {\sf jmason@flash.uchicago.edu; cattaneo@flash.uchicago.edu}}
\author{Stanislav Boldyrev}
\affiliation{Department of Physics, University of Wisconsin at Madison, 1150 University Ave, 
Madison, WI 53706; {\sf boldyrev@wisc.edu}}
\date{\today}

\begin{abstract}
We report the results 
of an extensive set of direct numerical simulations of forced,
incompressible,  
magnetohydrodynamic turbulence with a strong guide field. The aim is to
resolve the controversy regarding the power law exponent ($\alpha$, say) of the field
perpendicular energy 
spectrum $E(k_\perp) \propto k_\perp ^ {\alpha}$. The two main theoretical 
predictions,~$\alpha=-3/2$ and~$\alpha=-5/3$, have both received
 some support from numerical simulations carried out by different groups.   
Our simulations have a resolution 
of~$512^3$ mesh points, a strong guide field, an anisotropic simulation
domain, and implement 
a broad range of large-scale forcing routines, including those previously 
reported in the literature. Our findings indicate 
that the spectrum of well developed, strong incompressible MHD 
turbulence with a strong guide field is~$E(k_{\perp})\propto
k_{\perp}^{-3/2}$. 
\end{abstract}

\maketitle

{\em Introduction.}---Magnetohydrodynamic (MHD) turbulence is pervasive in astrophysical systems and plays an important 
role in stellar interiors, the solar wind and interstellar and intergalactic media (see \citep{biskamp}). 
Phenomenological models of 
MHD turbulence aim to describe the dynamics of energy 
transfer in the spectral domain. They are built upon the foundation that the spatial scales can be separated into three regions: 
(i) The energy containing range at large scales (small $k$) where energy is supplied to the system 
by an instability or an external force, (ii) the dissipation range at 
small scales (large $k$) where energy is removed from the system 
by viscosity or resistivity, 
and (iii) an intermediate region known as the inertial range. 
Within the inertial range it is assumed that forcing and dissipation are negligible and that energy is transferred from large to small  
scales solely by nonlinear interactions. It is therefore believed that, regardless of the form of the large-scale energy injection 
mechanism, once energy has cascaded to sufficiently small scales the nonlinear dynamics are universal 
(see, e.g., \citep{biskamp}).

Theoretical predictions for the scaling of the energy spectrum within the inertial range depend upon the 
assumed physics of the nonlinear interactions. The main theories assume that the basic state of MHD turbulence is one 
of Alfv\'en fluctuations: small-scale wave-packets, or eddies, propagate along the 
large-scale magnetic field with the Alfv\'en speed $V_A=B_0/\sqrt{4 \pi \rho}$, where $B_0$ is the magnitude of 
the large-scale field and $\rho$ is 
the fluid density. Only eddies propagating in opposite directions interact.
Iroshnikov \citep{iroshnikov} and Kraichnan \citep{kraichnan} used this fact to develop a theory for 
three-dimensionally isotropic eddies which need to undergo a large number of interactions to transfer 
energy to smaller scales. 
This leads to the Iroshnikov-Kraichnan energy spectrum $E(k) \propto k^{-3/2}$.

In later years, observational and numerical evidence revealed that the presence of a strong guiding magnetic field
 renders the turbulence anisotropic \citep{Shebalin,Strauss,Montgomery,Kadomtsev,Goldreich94}. 
This motivated Goldreich \& Sridhar \citep{Goldreich} to develop a new theory 
 in which 
the eddies are elongated in the direction of the large-scale field and the energy cascade proceeds mainly in the field 
perpendicular plane. The eddies are deformed strongly during only one interaction, leading to
the field-perpendicular energy spectrum $E(k_\perp) \propto k_\perp ^ {-5/3}$.
 
Recent high resolution numerical simulations verified the anisotropy of the turbulent cascade but also 
produced the Iroshnikov-Kraichnan exponent for the field-perpendicular 
spectrum \citep{marong2001,mullerbg2003,mullerg2005}, thus contradicting both the IK and GS models.  
To address this discrepancy a new theory was presented in \citep{boldyrev,boldyrev2}. 
In addition to the 
elongation of the eddies in the direction of the guiding field, therein it is proposed that the fluctuating velocity and magnetic 
fields at a scale $\lambda\sim 1/k_{\perp}$
are aligned within a small scale-dependent angle in the field perpendicular plane, $\theta \propto \lambda ^{1/4}$. The
process, known as scale-dependent dynamic alignment, reduces the strength of the nonlinear interactions 
and leads to the field-perpendicular energy spectrum  $E(k_{\perp}) \propto k_{\perp}^{-3/2}$.

Advances in computational power within recent years have made it feasible for 
a number of  independently working groups to test the above predictions 
via high resolution numerical simulations. The results have 
led to considerable debate. Although it is largely agreed upon that the turbulent cascade is anisotropic 
\citep[e.g.,][]{Shebalin,Strauss,Montgomery,Kadomtsev,Goldreich94,Goldreich,Bhattacharjee,Galtier,Chandran}, 
and excellent agreement with the scale-dependent alignment $\theta \propto \lambda ^{1/4}$ has been 
found \citep{mason}, 
there remains issues with regard to the exponent of the field-perpendicular energy 
spectrum. Some simulations have found $-5/3$ \cite{mullerb2000,haugen,haugen2,chov2000,cholv2002}, however, they did 
not have a strong guide field. Others yield $-3/2$, however, either their 
resolution was  
limited \cite{marong2001}, or 
the simulation domain was not anisotropic~\cite{mullerbg2003,mullerg2005}, which raised questions whether the field-parallel 
dynamics were captured. 
In addition, direct comparison of these numerical results is  
complicated by the fact that each have employed different forcing routines, different dissipation 
mechanisms, and have different Reynolds numbers.    

This principal controversy motivated our research. In an attempt to clarify the above issues, 
we have conducted a wide range of direct numerical simulations with a resolution of $512^3$ mesh points and 
Reynolds number~$Re\approx 2200$. 
Our simulations 
have a strong guide field, and an anisotropic simulation box to 
allow for the field-parallel dynamics. We force wavenumbers~$k=1,\,2$ and we have analyzed 17 different cases
in which we vary the relative intensities and correlation times of the forcing for the velocity and 
magnetic fields. These
cases include those settings previously reported in the literature. 

In almost all cases our simulations yield the spectrum~$-3/2$. A steeper spectrum,
consistent with~$-5/3$, is observed in a special case when the velocity field is driven by a force 
whose correlation time is much shorter than the relaxation time of the large-scale eddies. However,
we believe that this is a result of a setting that is not well suited for simulating well-developed strong 
MHD turbulence, since such short correlation time forcing may spoil an inertial interval of limited extent. 
In this case we do expect that the universal spectrum~$-3/2$ should emerge at smaller scales, though a larger 
calculation with a deeper inertial range is needed to observe it.

A further important finding 
is that in all cases the scale-dependent dynamic alignment of magnetic 
and velocity fluctuations, which is thought to be responsible for the~$-3/2$ 
spectrum \cite{boldyrev,boldyrev2,boldyrev3,mason}, is clearly 
observed. We therefore conclude that numerical simulations indicate 
that the spectrum of strong incompressible MHD 
turbulence with a strong guide field is~$E(k_{\perp})\propto k_{\perp}^{-3/2}$. 
In this paper we report the main results of our work. 
A detailed discussion will be presented elsewhere. \\

{\em Numerical results.}---We consider driven 
incompressible MHD turbulence with a strong guiding magnetic field. The equations read
\begin{eqnarray}
\label{eq:mhd}
{}&\partial_t {\bf v}+({\bf v}\cdot \nabla){\bf v}=-\nabla p+ \left(\nabla \times {\bf B}\right) \times {\bf B}+\nu \Delta 
{\bf v}+{\bf f_v},\nonumber \\
{}&\partial_t {\bf B} = \nabla \times ({\bf v} \times {\bf B})+\eta \Delta {\bf B}+{\bf f_B},
\end{eqnarray}
where ${\bf v }({\bf x}, t)$ is the velocity field, ${\bf B }({\bf x}, t)$ the magnetic field, $p$ the pressure,  and 
$\nu$ and $\eta$ are the fluid viscosity and resistivity, respectively. They will be solved using standard 
pseudospectral methods.  The external forces ${\bf f_{v}}({\bf x},t)$ 
and  ${\bf f_{B}}({\bf x},t)$ drive the turbulence at large scales, though the precise spatial and 
temporal form of the forcing is free to be chosen. In an attempt to resolve the previously reported 
controversies, we consider forcing mechanisms in the following 
three categories (collectively, they contain the main forcing mechanisms used in the recent literature).

{\bf Case 1:} Random forcing of the velocity field only (${\bf f_{B}}\equiv 0$) (e.g. \citep{chov2000,cholv2002}).

{\bf Case 2:} Random independent forcing of both the velocity and magnetic fields (e.g. \citep{marong2001}).

{\bf Case 3:} Steady forcing of both the velocity and magnetic fields by freezing large-scale modes (e.g. \citep{mullerg2005,mullerbg2003}). 

In cases 1 and 2, our random force satisfies the following requirements: it has 
no component along $z$, it is solenoidal in the $x-y$ plane, all the Fourier coefficients outside the range $1 \leq k \leq 2$ 
are zero, the Fourier coefficients inside that range are Gaussian random numbers with variance chosen so that 
the resulting {\it rms} velocity fluctuations are of order unity. The individual random values are refreshed 
independently {\it on average} at time intervals approximately twice as large as the turnover time of the large-scale eddies. We
also consider a special case with a renovation time that is ten times smaller.

In case 3 we initialize the calculation with a constant multiple of the statistically steady state solution from a simulation 
in which only the velocity is driven. 
We then evolve only those modes with $k \ge 2$ and we hold fixed the Fourier coefficients of both the velocity 
and the magnetic 
fields for those modes in which $k_x$, $k_y$ and $k_z$ are equal to $0$ or $\pm 1$ (excluding $k_x=k_y=k_z=0$). 
The multiplication factor is chosen 
so that the solution relaxes to a statistically steady state with {\it rms} velocity fluctuations of order unity.

The following results correspond to simulations that have an external magnetic field applied in the $z$ direction with
strength $B_0\approx 5$, measured in units of velocity. The periodic domain 
is elongated in the $z$ direction with aspect ratio 1:1:$B_0$. The Reynolds number is defined as 
$Re=U_{rms}(L/2\pi)/\nu$, where $L$ $(=2\pi)$ is the field-perpendicular box size, $\nu$ is fluid viscosity,
and $U_{rms}$ $(\sim 1)$ is the rms value of velocity fluctuations. We restrict ourselves to the case in which the magnetic 
resistivity and fluid viscosity are the same, $\nu=\eta$, with $Re\approx 2200$. The system is evolved until a stationary 
state is reached (confirmed by observing the time evolution of the total energy of fluctuations). The data sets for each run consist of 
approximately 30 samples that cover approximately 6 turnover times at the largest scales.

For each simulation we measure the two-dimensional energy spectrum, defined as 
$E(k_{\perp})=\langle |{\bf v}(k_{\perp})|^2\rangle k_{\perp} +\langle|{\bf b}(k_{\perp})|^2\rangle k_{\perp}$, 
where ${\bf v}(k_{\perp})$ and ${\bf b}(k_{\perp})$ are the two-dimensional Fourier transformations 
of the velocity and magnetic field in a plane perpendicular to ${\bf B}_0$ and $k_\perp=\left(k_x^2+k_y^2 \right)^{1/2}$. 
The average is taken over all such planes in the data cube, and then over all data cubes.  

\begin{figure} [tbp]
\vskip-10mm
\resizebox{0.5\textwidth}{!}{\includegraphics{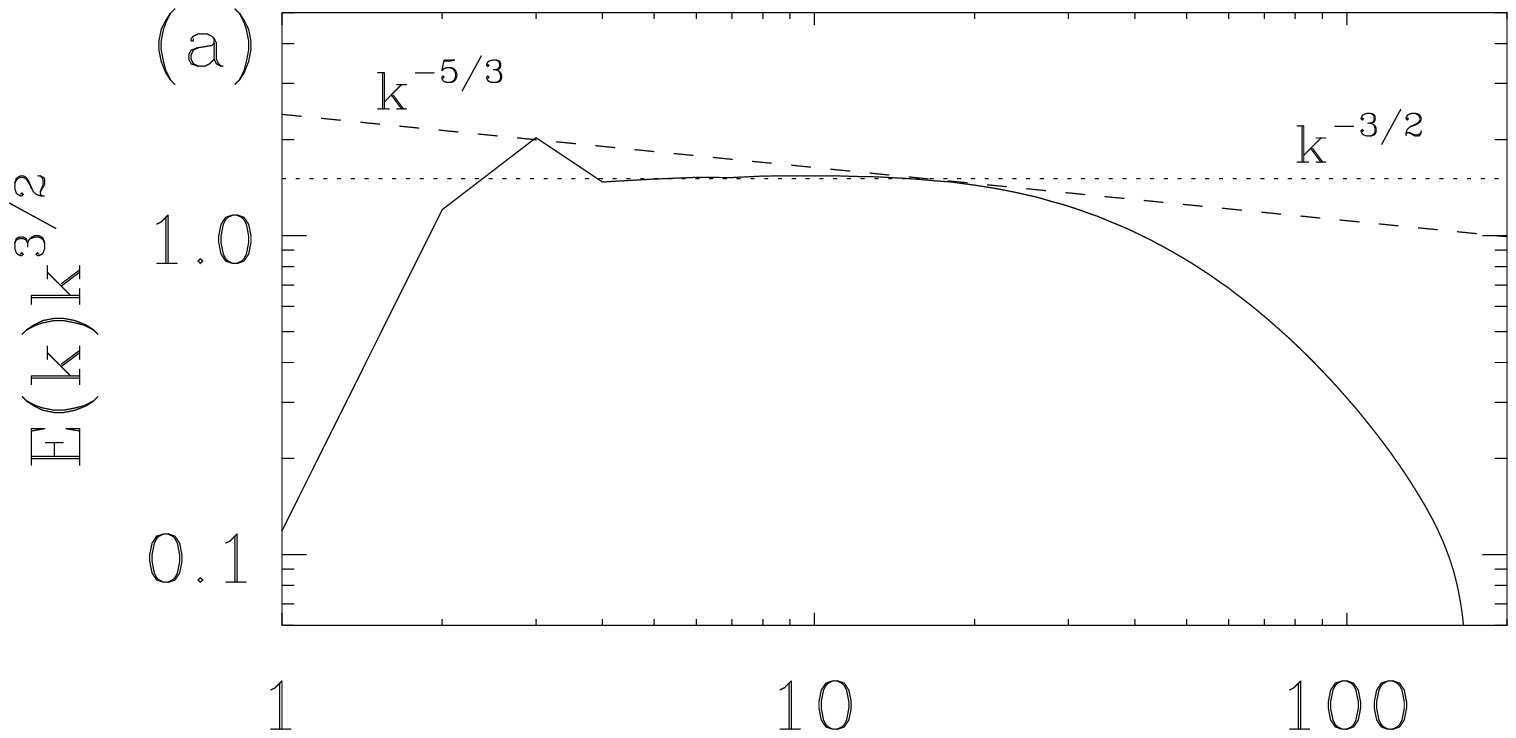}}
\vskip-18mm
\resizebox{0.5\textwidth}{!}{\includegraphics{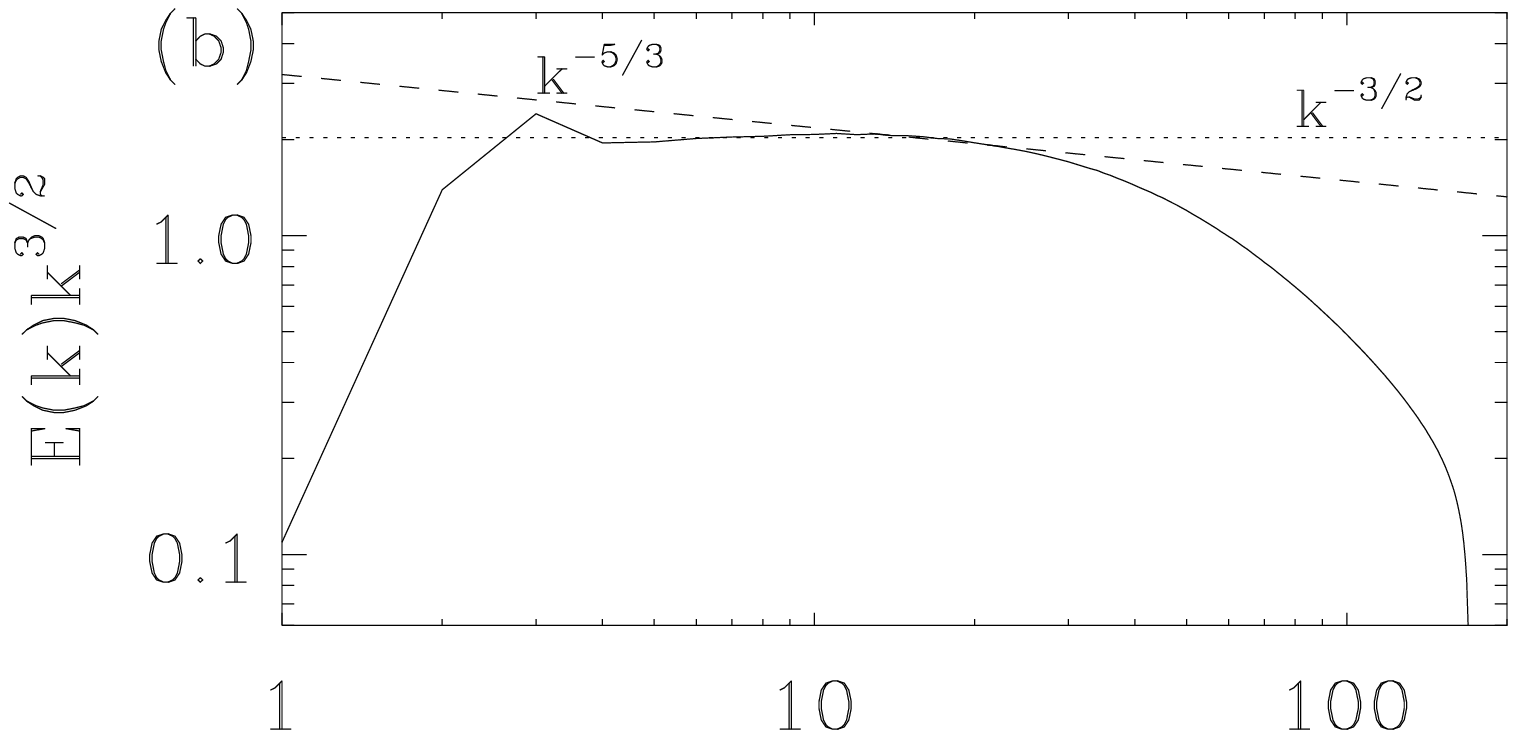}}
\vskip-18mm
\resizebox{0.5\textwidth}{!}{\includegraphics{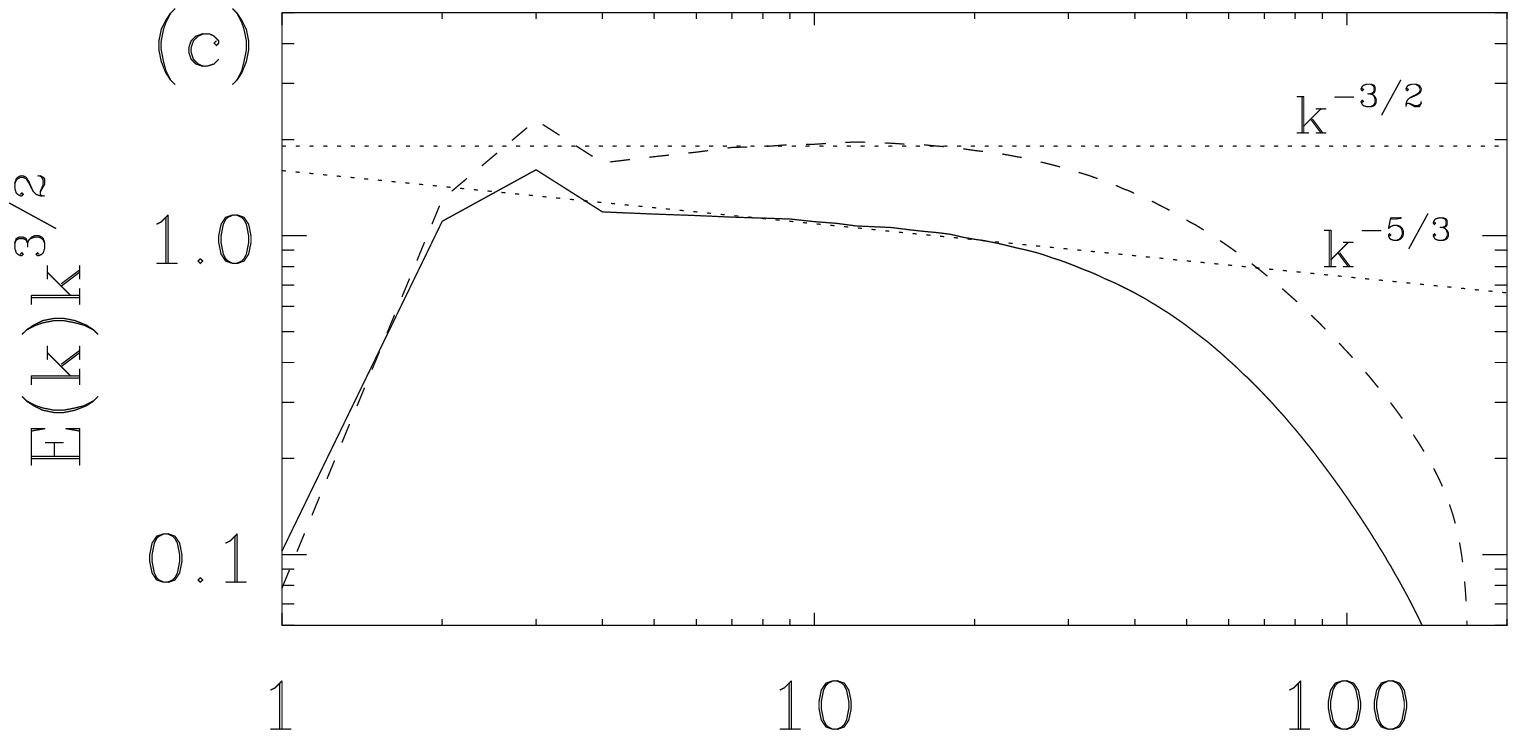}}
\vskip-18mm
\resizebox{0.5\textwidth}{!}{\includegraphics{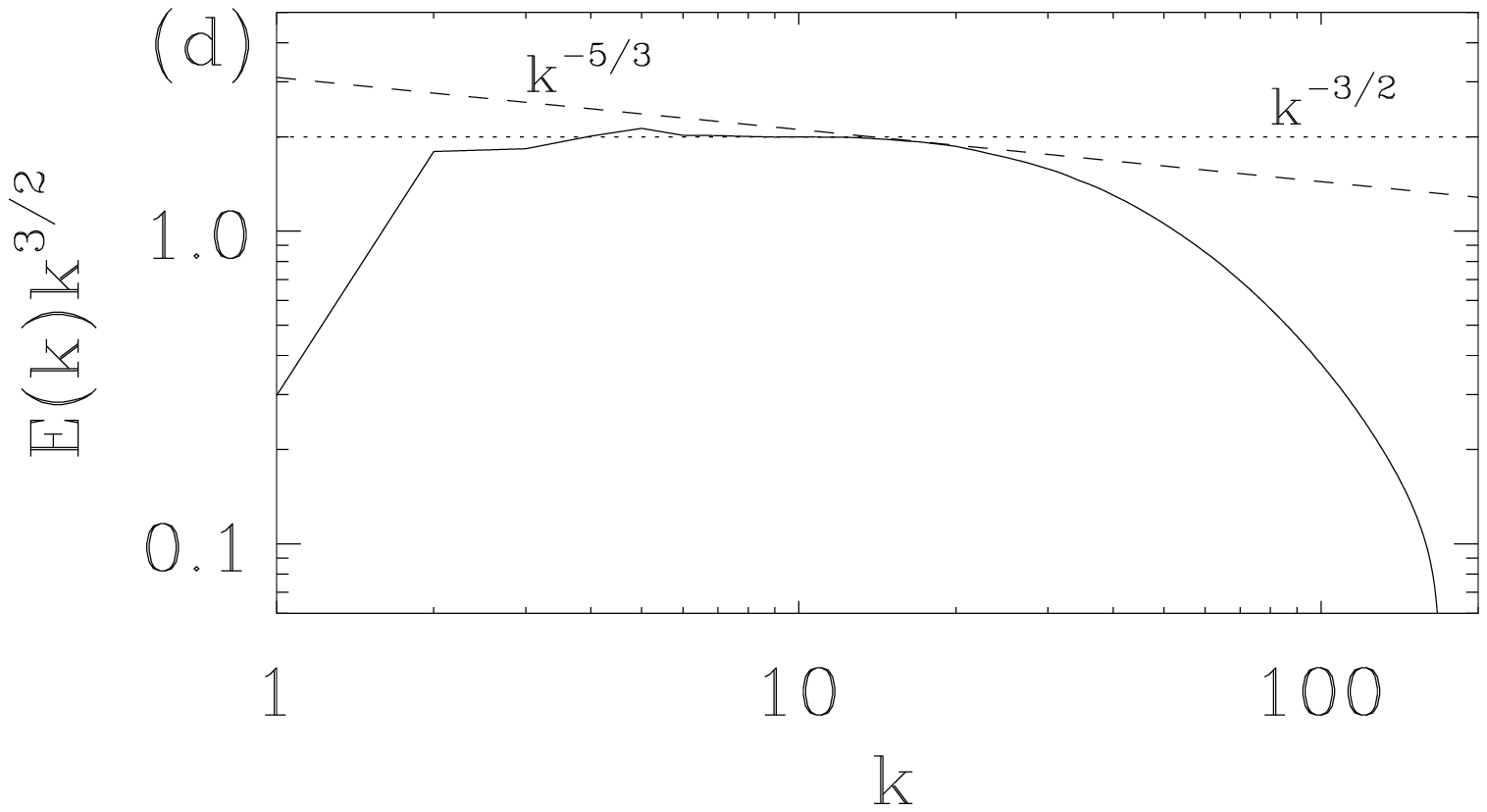}}
\vskip-8mm
\caption{The (compensated) field-perpendicular energy spectra $E(k_{\perp})$: (a) Case 1; (b) Case 2; 
(c) Case 1a (solid curve) and 2a (dashed curve); (d) Case 3. 
The description of the runs is given in Table~I. }
\label{spectrum}
\vskip-5mm
\end{figure} 

 Shown in Figure~\ref{spectrum}a is the spectrum of fluctuations for Case~1. 
 To infer whether a log-log plot of  $E(k_{\perp})$ has a slope of $-3/2$ or $-5/3$ we compensate the spectrum 
 (solid curve) and, for comparison,
  a function with scaling $k_{\perp}^{-5/3}$ (dashed line) by $k_{\perp}^{3/2}$. We conclude that the
best fit is $E(k_{\perp}) \propto k_{\perp}^{-3/2}$, with the inertial range corresponding to wavenumbers
$4 \lesssim k_{\perp} \lesssim 20$.  

The corresponding results for Case~2 are illustrated in
Figure~\ref{spectrum}b. Again the exponent $-3/2$ is a  better fit than $-5/3$, with the inertial range 
corresponding to wavenumbers $4 \lesssim k_{\perp} \lesssim 20$. 

In Figure~\ref{spectrum}c we show the spectra for the above two cases in a special setting when 
the forcing correlation time is made 10 times {\em shorter}. We find that when only the 
velocity field is forced (Case~1a: solid line) the best fit  
now appears to be~$-5/3$ with the inertial range corresponding to wavenumbers 
$5 \lesssim k_{\perp} \lesssim 23$. When both the magnetic and velocity fields are independently 
driven (Case~2a: dashed line)
the better fitting exponent remains at $-3/2$, although the range of wavenumbers over which this holds 
has become smaller and the fit is worse.

The energy spectrum for Case~3 is shown in Figure~\ref{spectrum}d. We note here that we have initialized 
the simulation
with the statistically steady state solution from Case~1a, for which we 
found the exponent~$-5/3$.  After a certain relaxation time, the energy spectrum of the 
resulting stationary turbulence is again better fit by~$-3/2$ with the inertial range corresponding 
to wavenumbers  $4 \lesssim k_{\perp} \lesssim 20$. The scaling exponents found in each case are summarized in Table~\ref{table1}. 
 In the next section we offer an explanation for our results. 

\begin{table}
\caption{Summary of the results. The forcing correlation time, $\tau$, is given in units of $(L/2\pi)/U_{rms}$.}
\vskip2mm
\begin{tabular}{|c|c|c|c|c|c|}
\hline
Case  & Forcing  & Forcing correlation time & Spectrum $E(k_\perp)$\\
\hline
1  & u only & 10 & $3/2$\\
\hline
2  & u and b & 10 & $3/2$\\
\hline
1a  & u only & 1 & $5/3$\\
\hline
2a  & u and b & 1 & $\lesssim 3/2$\\
\hline
3  & u and b & frozen large-scale modes & $3/2$\\
\hline
\end{tabular}
\label{table1}
\vskip-2mm 
\end{table}

\begin{figure} [tp]
\resizebox{0.5\textwidth}{!}{\includegraphics{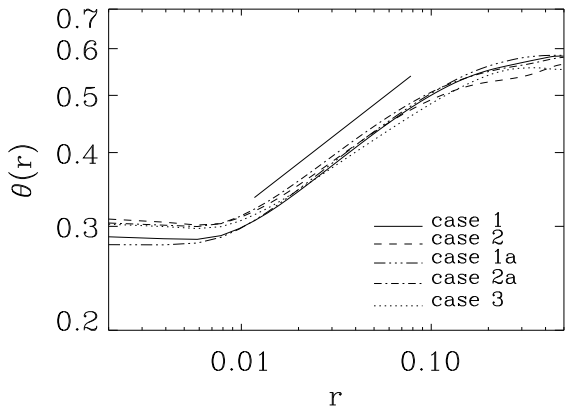}}
\caption{The alignment angle $\theta$ as a function of scale for each of the simulations. The straight  
line has slope 0.25 corresponding to the energy spectrum $E(k_{\perp})\propto k_{\perp}^{-3/2}$, 
see \citep{boldyrev,boldyrev2,mason}.}
\label{fig:angle}
\vskip-5mm 
\end{figure}

Finally, we also calculated the alignment angle 
between the shear-Alfv\'en velocity and magnetic field fluctuations. It is recalled that the theoretical 
prediction $\theta \propto \lambda^{1/4}$ is thought to be responsible for the field-perpendicular 
spectrum $E(k_{\perp})\propto k_{\perp}^{-3/2}$ \cite{boldyrev,boldyrev2,mason,boldyrev3}. 
 We define  
$\delta {\bf v}_{r}={\bf v}({\bf x}+{\bf r})-{\bf v}({\bf x})$ and $\delta {\bf b}_{r}={\bf b}({\bf x}+{\bf r})-{\bf b}({\bf x})$, 
where ${\bf r}$ is a  point-separation vector in a plane perpendicular to the large-scale field ${\bf B}_0$, and for each simulation 
we measure the ratio (see \citep{mason}) 
\begin{eqnarray}
\theta_{r} \approx \sin \left(\theta_r \right) \equiv \frac{\langle \vert \delta {\tilde {\bf v}}_{r}\times \delta {\tilde{\bf b}}_{r} \vert \rangle}
{\langle \vert \delta {\tilde{\bf v}}_{r} \vert \vert  \delta {\tilde{\bf b}}_{r} \vert \rangle},
\label{angle}
\end{eqnarray} 
where $\delta {\tilde {\bf v}}_{r}=\delta {\bf v}_{r}-(\delta 
{\bf v}_{r}\cdot {\bf n}){\bf n}$, $\delta {\tilde {\bf b}}_{r}=\delta {\bf b}_{r}-(\delta {\bf b}_{r}\cdot {\bf n}){\bf n}$
 and ${\bf n}={\bf B}({\bf x})/\vert {\bf B}({\bf x})\vert$. Figure~\ref{fig:angle} 
shows that $\theta \propto \lambda^{1/4}$ is good fit for {\it all} the simulations. 
We note that the Goldreich \& Sridhar theory 
\citep{Goldreich} which predicts $E(k_{\perp})\propto k_{\perp}^{-5/3}$ would be consistent with a zero slope in 
Figure~\ref{fig:angle}. We propose that the reason that the alignment angle
is well observed is because its measurement is composed of the ratio of two structure functions, and as
such is somewhat analogous to the phenomenon of extended self-similarity~\citep{Benzi}. This phenomenon is well known 
in hydrodynamic turbulence, where a cleaner scaling behavior and longer inertial ranges are apparent 
when one structure function is considered as a function of the other, rather than considering each separately 
as functions of~$r$. \\

{\em Discussion and conclusion.}---
Our results lead us to believe that the universal spectrum for MHD turbulence in the presence of
a strong guiding field is $E(k_{\perp})\propto k_{\perp}^{-3/2}$. 
In order to understand the exceptional case (Case~1a) in which we obtain
an exponent closer to $-5/3$, it is instructive to 
determine the turnover time of the large-scale eddies. This can be done by measuring the energy relaxation time in Case~3, 
i.e., the characteristic time on which the solution adjusts from a randomly stirred state to a 
state with frozen large-scale modes. We find that the large-scale eddy turnover time is approximately~$\tau_0\sim 5$, which 
agrees with the dimensional estimate~$\tau_0 \approx L/U_{rms}\approx 2 \pi$. 

We then conducted a series of simulations 
(not shown here) analogous to Case~1a where we varied the force 
correlation time from $\tau=0.1$ to $\tau=10$. We observed that the transition from 
the spectrum $E(k_{\perp})\propto k_{\perp}^{-5/3}$ to $E(k_{\perp})\propto k_{\perp}^{-3/2}$ occurred  
between $\tau \approx 1$ and $\tau\approx 2$, which is much smaller than the large-scale eddy turnover 
time. We believe that a driving mechanism with such a short renovation time is not well suited for investigations 
of well developed strong MHD turbulence, since an inertial interval of limited extent may be spoiled by 
the transition region from 
the forcing interval at scales where the eddy turnover times are larger than the force correlation time. We believe it
is this feature that may be responsible for the different exponent in this case. It is reasonable to expect that 
a larger calculation would yield 
$E(k_{\perp})\propto k_{\perp}^{-3/2}$ sufficiently deep in the inertial range, though at the 
present time such a calculation is not feasible. This issue is currently under investigation by different means, 
and will be addressed in detail elsewhere.

\acknowledgments
We thank Jean C. Perez for useful discussions. This work was supported by the NSF Center for Magnetic
Self-Organization in Laboratory and Astrophysical Plasmas
at the University of Chicago and 
the University of Wisconsin at Madison. We appreciate the hospitality 
and support of the Aspen Center for Physics where a part of this work 
was performed. We gratefully acknowledge use of ``BGL", 
a 1024-node IBM Blue Gene/L system operated by the Argonne Leadership 
Computing facility at Argonne National Laboratory.


\begin{thebibliography}{99}
\bibitem[Biskamp(2003)]{biskamp} Biskamp, D., 2003, {\em Magnetohydrodynamic Turbulence.} (Cambridge University Press, Cambridge).
\bibitem[Iroshnikov(1963)]{iroshnikov} Iroshnikov, P. S.,  AZh, {\bf 40} (1963) 742; Sov. Astron, {\bf 7} (1964) 566.
\bibitem[Kraichnan(1965)]{kraichnan} Kraichnan, R. H.,  Phys. Fluids, {\bf 8} (1965) 1385.
\bibitem[Kadomtsev(1974)]{Kadomtsev} Kadomtsev, B.~B. \& Pogutse, O.~P. Sov. Phys.- JETP, {\bf 38} (1974) 283.
\bibitem[Strauss(1976)]{Strauss} Strauss, H.~R., Phys. Fluids, {\bf 19} (1976) 134.
\bibitem[Montgomery(1982)]{Montgomery} Montgomery, D., Phys. Scripta, {\bf T2} (1982) 83.
\bibitem[Shebalin(1983)]{Shebalin} Shebalin, J.~V., Matthaeus, W.~H. \& Montgomery, D. J.~Plasma Physics, 
{\bf 29} (1983) 525.
\bibitem[Sridhar \& Goldreich(1994)]{Goldreich94} Sridhar, S. \& Goldreich, P. ApJ {\bf 432} (1994) 612.
\bibitem[Goldreich \& Sridhar(1995)]{Goldreich} Goldreich, P. \& Sridhar, S., ApJ {\bf 438} (1995) 763.
\bibitem[Maron \& Goldreich(2001)]{marong2001} Maron, J., \& Goldreich, P., ApJ {\bf 554} (2001) 1175.
\bibitem[M\"uller \& Grappin(2005)]{mullerg2005} M\"uller, W.-C. \& Grappin, R., Phys. Rev. Lett., {\bf 95} (2005) 114502.
\bibitem[M\"uller, Biskamp \& Grappin(2003)]{mullerbg2003}  M\"uller, W.-C., 
Biskamp, D., \& Grappin, R., Phys. Rev. E~{\bf 67} (2003) 066302.  
\bibitem[Boldyrev(2005a)]{boldyrev} Boldyrev, S., ApJ {\bf 626} (2005) L37.
\bibitem[Boldyrev(2005b)]{boldyrev2} Boldyrev, S., Phys. Rev. Lett. {\bf 96} (2006) 115002; astro-ph/0511290.
\bibitem[Galtier, Pouquet \& Mangeney(2005)]{Galtier} Galtier, S., Pouquet, A.,  \&  Mangeney, A., Phys. Plasmas {\bf 12} (2005) 092310. 
\bibitem[Chandran(2004)]{Chandran} Chandran, B. D. G., Astrophys. Space Sci. {\bf 292} (2004) 17.
\bibitem[Ng \& Bhattacharjee(1996)]{Bhattacharjee} Ng, C. S., \& Bhattacharjee, A., ApJ {\bf 465} (1996) 845. 
\bibitem[Mason, Cattaneo, Boldyrev(2006)]{mason} Mason, J., Cattaneo, F., \& Boldyrev, S., Phys. Rev. Lett, {\bf 97} (2006) 255002; astro-ph/0602382. 
\bibitem[M\"uller \& Biskamp(2000)]{mullerb2000} M\"uller, W.-C. \&  
Biskamp, D., Phys. Rev. Lett., {\bf 84} (2000) 475.  
\bibitem[Haugen, Brandenburg, \& Dobler(2004a)]{haugen} Haugen, N. E. L., Brandenburg, A., \& Dobler, W., Phys. Rev. E~{\bf 70} (2004) 016308. 
\bibitem[Haugen, Brandenburg, \& Dobler(2004b)]{haugen2} Haugen, N. E. L., Brandenburg, A., \& Dobler, W., Astrophys. Space Sci. {\bf 292} (2004) 53. 
\bibitem[Cho \& Vishniac(2000)]{chov2000} Cho, J. \& Vishniac, E.~T., ApJ. {\bf 539} (2000) 273.
\bibitem[Cho, Lazarian \& Vishniac(2000)]{cholv2002} Cho, J., Lazarian, A., \& Vishniac, E.~T., ApJ. {\bf 564} (2002) 291.
\bibitem[Boldyrev, Mason \& Cattaneo(2006)]{boldyrev3} Boldyrev, S., Mason, J. \& Cattaneo, F., (2006) arXiv:astro-ph/0605233. 
\bibitem[Benzi et al.(1993)]{Benzi} Benzi, R., Ciliberto, S., Tripiccione, R., Baudet, C., Massaioli, F., and Succi, S.,
   Phys. Rev. E {\bf 48} (1993) R29.

\end{thebibliography}
\end{document}